\documentclass{article}
\usepackage[utf8]{inputenc}

\usepackage{graphicx}
\usepackage{amsmath}
\usepackage{amsfonts}
\usepackage{amssymb}
\usepackage{bm}
\usepackage{graphicx}
\usepackage{xcolor}
\usepackage{subcaption}
\usepackage{booktabs}
\usepackage[ruled,vlined]{algorithm2e}
\usepackage{subcaption}
\usepackage{natbib}
\SetKwInput{KwInput}{Input}                
\SetKwInput{KwOutput}{Output}              

\newcommand{\X}{\boldsymbol{X}}

\newcommand{\R}{\boldsymbol{R}}
\newcommand{\Y}{\boldsymbol{Y}}

\newcommand{\e}{\boldsymbol{e}}

\newcommand{\SIGMA}{\boldsymbol{\Sigma}}

\newcommand{\PSI}{\bm{\Psi}}

\newcommand{\GAMMA}{\bm{\Gamma}}

\newcommand{\re}[1]{\textcolor{black}{#1}} 

\DeclareMathOperator*{\tr}{tr}
\DeclareMathOperator*{\argmin}{arg\,min}
\DeclareMathOperator*{\vect}{vec}

\title{A general framework for penalized mixed-effects multitask learning with applications on DNA methylation surrogate biomarkers creation}

\author{Andrea Cappozzo \footnote{MOX Lab, Department of Mathematics, Politecnico di Milano, \texttt{andrea.cappozzo@polimi.it, francesca.ieva@polimi.it}}         \and
        Francesca Ieva\footnotemark[\value{footnote}] \footnote{Health Data Science Center, Human Technopole} \and
        Giovanni Fiorito \footnote{Laboratory of Biostatistics, Department of Biomedical Sciences, University of Sassari \texttt{gfiorito@uniss.it}}
}

\begin{document}
\date{\vspace{-5ex}}
\maketitle

\begin{abstract}
Recent evidence highlights the usefulness of DNA methylation (DNAm) biomarkers as surrogates for exposure to risk factors for non-communicable diseases in epidemiological studies and randomized trials. DNAm variability has been demonstrated to be tightly related to lifestyle behavior and exposure to environmental risk factors, ultimately providing an unbiased proxy of an individual state of health. At present, the creation of DNAm surrogates relies on univariate penalized regression models, with elastic-net regularizer being the gold standard when accomplishing the task. Nonetheless, more advanced modeling procedures are required in the presence of multivariate outcomes with a structured dependence pattern among the study samples. In this work we propose a general framework for mixed-effects multitask learning in presence of high-dimensional predictors to develop a multivariate DNAm biomarker from a multi-center study. A penalized estimation scheme based on an expectation-maximization algorithm is devised, in which any penalty criteria for fixed-effects models can be conveniently incorporated in the fitting process. We apply the proposed methodology to create novel DNAm surrogate biomarkers for multiple correlated risk factors for cardiovascular diseases and comorbidities. We show that the proposed approach, modeling multiple outcomes together, outperforms state-of-the-art alternatives, both in predictive power and bio-molecular interpretation of the results.
\end{abstract}
\noindent
\section{Introduction}
DNA methylation (DNAm) is an epigenetic process that regulates gene expression, typically occurring in cytosine within CpG sites (CpGs) in the DNA sequence \citep{Singal1999}. \re{DNAm regulates gene expression in different manners. Specifically, high DNAm has been observed in bodies of highly transcribed genes, whereas DNAm in gene promoters and first introns typically have an inverse correlation with gene expression \citep{Anastasiadi2018,Rauluseviciute2020}. Also, recent studies suggest that the relationship between genetic variation, DNAm and gene expression is complex and tissue-specific, highlighting that DNAm in non-CpG island regions regulates the transcription of distal genes \citep{VanEijk2012}. Advanced technology allows measuring whole-genome DNAm for many samples at the same time. The most common ways for DNAm measurements consist of whole-genome bisulphite sequencing and DNAm microarray. The first commercial high-density microarray measuring genome-wide methylation was the HumanMethylation27 (27K CpGs) released by Illumina in 2009, followed by the HumanMethylation450 (450K CpGs), and more recently, by the IlluminaMethylation850 \citep[850K CpGs,][]{Campagna2021}. Since then, a tremendous amount of associations between DNAm at individual CpG sites and different exposures, traits, and diseases have been identified in the so-called epigenome-wide association studies \citep[EWAS,][]{Battram2021}}. 
Concurrently, the development of surrogate scores based on blood DNA methylation has also received thriving attention in recent years: impressive epidemiological evidence has been established between DNAm and individual history of exposure to lifestyle and environmental risk factors \citep{Zhong2016, Guida2015, Fiorito2018}. To this extent, multi-CpG DNAm biomarkers have been devised to predict patient-specific state of health indicators; and relevant examples include epigenetic clocks to measure ``biological age'' \citep{Lu2019}, smoking habits \citep{Guida2015} and proxies for inflammatory proteins \citep{Stevenson2020}. Remarkably, DNAm based scores have been demonstrated to outperform surveyed exposure measurements when  predicting diseases \citep{Zhang2016, Conole2020}. A possible explanation for this somewhat counter-intuitive behavior being that DNA methylation intrinsically accounts for biases in self-reported exposure (e.g., underestimation of smoked cigarettes) as well as individual responses to risk factors (e.g., the same amount of tobacco may produce different effects in dissimilar patients).


From a modeling perspective, state-of-the-art methods for DNAm biomarkers creation generally rely on standard univariate penalized regression models, with elastic-net \citep{Zou2005} being the routinely employed technique when accomplishing the task. Indeed, the associated learning problem entirely falls within the ``$p$ bigger than $N$'' framework: DNA methylation levels are measured at approximately half million CpG sites for each sample, with the dimension of the latter generally not exceeding the order of thousands in most studies. The afore-described procedure is shown to be widely effective in building DNAm biomarkers, with very recent contributions including surrogate scores for short-term risk of cardiovascular events \citep{Cappozzo2022}, cumulative lead exposure \citep{Colicino2021}, DNAm surrogate for alcohol consumption, obesity indexes, and blood measured inflammatory proteins \citep{Marioni2021}  and the identification of CpG sites associated with clinical severity of COVID-19 disease \citep{CastrodeMoura2021}. 
Nonetheless, elastic-net penalties may be too restrictive when dealing with complex learning problems involving multivariate responses and distinctive dependence patterns across statistical units. 

The aforesaid first layer of complexity is encountered when a multi-dimensional DNAm biomarker needs to be created, to jointly model multiple risk factors and to coherently account for the correlation structure among the response variables. Such a multivariate problem, also known as multi-task regression in the machine learning literature \citep{Caruana1997}, can be fruitfully untangled only if dedicated care is devoted in choosing the most appropriate penalty required for the analysis. For instance, one may opt for the incorporation of  $\ell1/\ell2$ type of regularizers \citep{Obozinski2009,Obozinski2010, Li2015}, that extend the lasso \citep{Tibshirani1996}, group-lasso \citep{Yuan2016} and sparse group-lasso \citep{Simon2013, Laria2019} to the multiple response framework. Another option could contemplate the inclusion, within the estimation procedure, of prior information related to the association structure among CpG sites: this is  effectively achieved by means of graph-based penalties \citep{Li2010, Kim2013, Cheng2014, Dirmeier2018}. Furthermore, tree-based regularization methods have also been recently introduced in the literature, to account for hierarchical structure over the responses in a single study \citep{Kim2012} as well as when multiple data sources are at our disposal \citep{Zhao2020, Zhao2022}. For a thorough and up-to-date survey on the analysis of high-dimensional omics data via structured regularization we refer the interested reader to \cite{Vinga2021}, while the monograph of \cite{Hastie2015} provides a general introduction to statistical learning with sparsity.

A second layer of complexity is introduced when DNA samples and related blood measured biomarkers are collected in a study comprising multiple cohorts. In such a situation, an unknown degree of heterogeneity may be included in the data, with patients coming from the same cohort sharing some degree of commonality. Observations in the dataset are thus no longer independent and the cohort-wise covariance structure needs to be properly estimated. Linear Mixed-Effects Models (LMM) provide a convenient solution to this problem by adding a random component to the model specification \cite[see, e.g.,][for an introduction on the topic]{Pinheiro2006, Galecki2013, Demidenko2013}. Whilst being able to capture unobserved heterogeneity, standard mixed models, very much like their fixed counterpart, cannot directly handle situations in which the number of predictors exceeds the sample size. In order to overcome this issue \cite{Schelldorfer2011} introduced a procedure for estimating high-dimensional LMM via an $\ell_1$-penalization 
.	More recently, \cite{Rohart2014} devised a general-purpose ECM algorithm \citep{Meng1993} for solving the same issue,  but achieving greater flexibility as the proposed framework can be combined with any penalty structure previously developed for linear fixed-effects models.

A  Multivariate Mixed-Effects Model (MLMM) is an LMM in which multiple characteristics (response variables) are measured for the statistical units comprising the study. Despite being quite a long-established methodology \citep{Reinsel1984, Shah1997}, its further development has not received much attention in the recent literature. Relevant exceptions include the computational strategies for handling missing values proposed in \cite{Schafer2002}, and the estimation theory based on hierarchical likelihood developed in \cite{Chipperfield2012}. On this account, to the best of our knowledge, a unified approach for penalized  MLMM estimation is still missing in the literature and it could thus be a relevant contribution to the statistics and machine learning fields.





Motivated by the problem of creating a DNAm biomarker for hypertension and hyperlipidemia from a multi-center study, we propose in this article a general framework for high-dimensional multitask learning with random-effects. Leveraging from the algorithm introduced in \cite{Rohart2014} for the univariate response case, the estimation mechanism is effectively constructed to accommodate custom penalty types, building upon existing routines developed for regression with fixed-effects only. 

The remainder of the paper is structured as follows. Section \ref{sec:EPIC_data} describes the EPIC Italy dataset, which gave the motivation for the development of the methodology proposed in this manuscript. In Section \ref{sec:methodology} we introduce the penalized mixed-effects model for multitask learning, covering its formulation, inference and model selection. Section \ref{sec:sim_study} presents a simulation study on synthetic data for three different scenarios. Section \ref{sec:application} outlines the results of the novel method applied to the EPIC Italy data for creating DNAm surrogates for cardiovascular risk factors and comorbidities, comparing it with state-of-the-art alternatives.  Section \ref{sec:conclusion} concludes the paper with a discussion and directions for future research. The \texttt{R} package \texttt{emlmm} implementing the proposed method accompanies the article and it is freely available at \texttt{https://github.com/AndreaCappozzo/emlmm}.

\section{EPIC Italy data and study design} \label{sec:EPIC_data}
The considered dataset belongs to the Italian branch of the European Prospective Investigation into Cancer and Nutrition (EPIC) study, one of the largest cohort study in the world, with participants recruited across 10 European countries and followed for almost 15 years \citep{Riboli2002}. For each participant, lifestyle and personal history questionnaires were recorded, together with anthropomorphic measures and blood samples for DNA extraction. The EPIC Italy dataset is comprised of geographical sub-cohorts identified by the center of recruitment; particularly, we will consider the provinces of Ragusa and Varese and the cities of Turin and Naples. The latter center became associated with EPIC in later times through the Progetto ATENA study \citep{Panico1992}. \re{DNAm was measured with the HumanMethylation450 array following standard laboratory procedures \citep[see][for a detailed description]{Fiorito2022}, while the pre-processing included removing CpG sites and samples with a call rate lower than 95\%, BMIQ method for reducing technical variability and bias introduced by type II probes, and ComBat technique for batch effect adjustment \citep{Marabita2013a}.}

By profiting from the information recorded in the aforementioned sub-cohorts, we aim at creating a multi-dimensional DNAm biomarker for cardiovascular risk factors and comorbidities. To this extent, we consider a multivariate  response comprised of $r=5$ measures, namely Diastolic Blood Pressure (DBP), Systolic Blood Pressure (SBP),  High Density Lipoprotein (HDL), Low Density Lipoprotein (LDL) and Triglycerides (TG). These characteristics are chosen as they represent the major risk factors for cardiovascular diseases \citep{Wu2015}. In building a DNAm biomarker, the response variables are regressed on DNA methylation values for each CpG site, adjusted for sex and age. A total of $N=574$ individuals in the $J=4$ cohorts showcase non-missing values for every response variable: they comprise the  sample onto which all subsequent analyses will be performed. \re{To reconstruct the process of DNAm surrogates creation and validation, the EPIC Italy data is randomly split into two sets: $70\%$ ($N_{tr}=401$ ) of it is employed for pre-processing and model fitting, while the remaining $30\%$ ($N_{te}=173$ ) acts as test set for assessing prediction accuracy. In addition, we will consider samples from the EXPOsOMICS project \citep{Fiorito2018} as an external validation dataset to assess out of groups predictive performance. In details, EXPOsOMICS is a case-control study on cardiovascular diseases (CVDs) nested in the EPIC Italy cohort composed by $276$ volunteers (not overlapping with the main dataset) whose center of recruitment is unknown or different from the $J=4$ observed in the learning phase.}

\re{Coming back to the data analysis pipeline, an epigenome-wide association study \citep[EWAS,][]{Campagna2021} is performed on the training set as a pre-screening procedure. In details, log-transformed DBP, SBP, HDL, LDL and TG are separately regressed on each available CpG site, adjusting for sex and age. P-values are then collected and arranged in increasing order. We then screen the set of predictors retaining, for each dimension of the multivariate response, the CpG sites whose p-values are smaller than the fifth percentile of the resulting empirical distributions. The final set of covariates for the multitask learning problem is achieved by taking the union of the resulting CpG sites separately preserved for DBP, SBP, HDL, LDL and TG.
In so doing, out of the whole initial set of $295614$ CpG sites, $62128$ DNA methylation features are retained for subsequent modeling. Together with sex and age, this amounts to a total of $p=62130$ predictors and a $5$-dimensional response for a training sample size of $N_{tr}=401$.} Whilst variable screening in ultra-high feature space is itself an ongoing research field \cite[see, e.g.,][and references therein]{Fan2008, Fan2009,Zhong2021}, we decided to rely on the EWAS technique as it is the standard approach employed in epigenomics \citep{Fazzari2010}. 

\begin{figure}
    \centering
    \includegraphics[width=\textwidth,keepaspectratio]{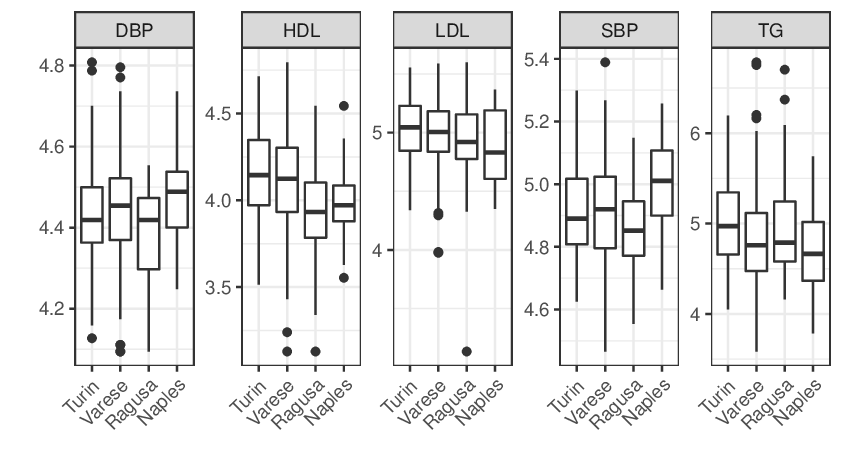}
    \caption{Boxplots of log-transformed Diastolic Blood Pressure (DBP), High Density Lipoprotein (HDL),  Low Density Lipoprotein (LDL), Systolic Blood Pressure (SBP)  and Triglycerides (TG) for different Center, Italy EPIC training dataset.}
    \label{fig:boxplot_Y}
\end{figure}

\begin{figure}
    \centering
    \includegraphics[scale=.9]{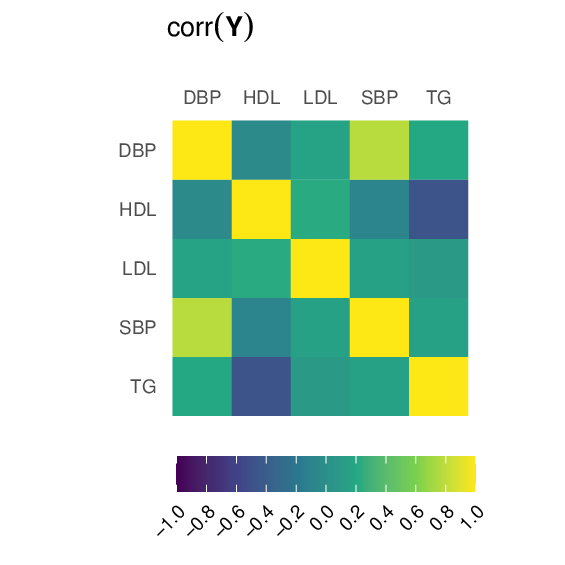}
    \caption{Sample correlation matrix of log-transformed Diastolic Blood Pressure (DBP), High Density Lipoprotein (HDL),  Low Density Lipoprotein (LDL), Systolic Blood Pressure (SBP)  and Triglycerides (TG), Italy EPIC training dataset.}
    \label{fig:corrplot_Y}
\end{figure}

As previously mentioned, the considered training samples belong to four different centers distributed across Italy, with data for $91$, $ 234$,   $ 44$   and  $32$ volunteers respectively collected in Turin, Varese, Ragusa and Naples provinces. The boxplots in Figure \ref{fig:boxplot_Y} emphasize the differences in the five response variables by center. To capture the center-wise variability and to maintain generalizability of the devised DNAm biomarker outside the Italy EPIC cohorts, a partial pooling random-intercept model shall be adopted. That is, a $q=1$ random-effect component is included in the model specification. Furthermore, the biomarkers comprising the response vector showcase some degree of relations,  as displayed by the sample correlation matrix of Figure \ref{fig:corrplot_Y}, so much so that it is sensible to regress them jointly to take advantage of their association structure in the model formulation.
This challenging learning task requires an ad-hoc specification for a multivariate mixed-effects framework applicable to high-dimensional predictors.

\section{Penalized mixed-effects model for multitask learning} \label{sec:methodology}
In this section, a novel approach for multivariate mixed-effects modeling based on penalized estimation is proposed.
\subsection{Model definition} \label{sec:mod_def}
The multivariate linear mixed-effects model \citep{Shah1997} expresses the $n_j \times r$ response matrix $\Y_j$ for the $j$-th group as:
\begin{equation} \label{eq:r_dim_lmm}
\boldsymbol{Y}_{j}=\boldsymbol{X}_{j} \boldsymbol{B}+\boldsymbol{Z}_{j} \boldsymbol{\Lambda}_{j}+\boldsymbol{E}_{j},
\end{equation}
where, for each of the $n_j$ samples in group $j$ and  $\sum_{j=1}^Jn_j=N$, $r$ response variables have been measured. The remainder terms define the following quantities:
\begin{itemize}
\item $\boldsymbol{B}$ is the $p\times r$ matrix of fixed-effects (including the intercept)
\item $\boldsymbol{\Lambda}_j$ is the $q \times r$ matrix of random-effects
\item $\boldsymbol{X}_{j}$ is the $n_{j} \times p$ fixed-effects design matrix 
\item $\boldsymbol{Z}_{j}$ is the $n_{j} \times q$ random-effects design matrix
\item $\boldsymbol{E}_{j}$ is the $n_{j} \times r$ within-group error matrix
\item $j=1,\ldots,J$, with $J$ total number of groups.
\end{itemize}
By employing the vec operator, we assume that:
\[\vect \left(\boldsymbol{\Lambda}_j \right)\sim \mathcal{N} (\boldsymbol{0},\boldsymbol{\Psi}),\]
where $\boldsymbol{\Psi}$ is a $qr \times qr$ positive semidefinite matrix, incorporating variations and covariations between the $r$ responses and the $q$ random-effects. We further assume that the error term is distributed as follows:
\begin{equation} \label{eq:E_term}
\vect \left(\boldsymbol{E}_j \right)\sim \mathcal{N} (\boldsymbol{0},\boldsymbol{\Sigma} \otimes \boldsymbol{I}_{n_j}),
\end{equation}
where $\boldsymbol{\Sigma}$ is a $r \times r$ covariance matrix, capturing dependence among responses, and $\boldsymbol{I}_{n_j}$ is the identity matrix of dimension $n_j \times n_j$. Formulation in \eqref{eq:E_term} explicitly induces independence between the row vectors of $\boldsymbol{E}_j$. Therefore, the entire model can be rewritten in vec form:
\[\vect \left(\boldsymbol{Y}_j \right) \sim N\left( (\boldsymbol{I}_{r} \otimes \boldsymbol{X}_{j}) \vect \left( \boldsymbol{B} \right), \left( \boldsymbol{I}_r \otimes \boldsymbol{Z}_{j}\right)\boldsymbol{\Psi}\left( \boldsymbol{I}_r \otimes \boldsymbol{Z}_{j}\right)^{'} + \boldsymbol{\Sigma} \otimes \boldsymbol{I}_{n_j}\right).\]
Given a sample of $N=\sum_{j=1}^J n_j$, the log-likelihood of model \eqref{eq:r_dim_lmm} reads:
\begin{align}\label{eq:log_lik}
\begin{split}
&\ell(\boldsymbol{\theta})=\sum_{j=1}^J -\frac{n_j}{2}\log{2 \pi}-\frac{1}{2}\log| \left( \boldsymbol{I}_r \otimes \boldsymbol{Z}_{j}\right)\boldsymbol{\Psi}\left( \boldsymbol{I}_r \otimes \boldsymbol{Z}_{j}\right)^{'} + \boldsymbol{\Sigma} \otimes \boldsymbol{I}_{n_j}|+\\
&-\frac{1}{2}\left(\vect \left(\boldsymbol{Y}_j \right)- (\boldsymbol{I}_{r} \otimes \boldsymbol{X}_{j}) \vect \left( \boldsymbol{B} \right) \right)^{'} \left(\left( \boldsymbol{I}_r \otimes \boldsymbol{Z}_{j}\right)\boldsymbol{\Psi}\left( \boldsymbol{I}_r \otimes \boldsymbol{Z}_{j}\right)^{'} + \boldsymbol{\Sigma} \otimes \boldsymbol{I}_{n_j}\right)^{-1} \left(\vect \left(\boldsymbol{Y}_j \right)- (\boldsymbol{I}_{r} \otimes \boldsymbol{X}_{j}) \vect \left( \boldsymbol{B} \right) \right),
\end{split}
\end{align}
where $\boldsymbol{\theta}=\{\boldsymbol{B}, \SIGMA,\PSI\}$ is the set of parameters to be estimated. When the framework outlined in \eqref{eq:r_dim_lmm} is employed for DNAm biomarker creation, the number of regressors $p$ is most certainly much larger than the sample size $N$. We are thus not directly interested in maximizing \eqref{eq:log_lik}, but rather a penalized version of it, generically defined as follows:
\begin{equation} \label{eq:pen_log_lik}
\ell_{pen}(\boldsymbol{\theta})=\ell(\boldsymbol{\theta})-p(\boldsymbol{B} ; \lambda),
\end{equation}
with $p(\boldsymbol{B} ; \lambda)$ being a penalty term employed to regularize the fixed-effects $\boldsymbol{B}$ as a function of the complexity parameter $\lambda \geq 0$. Notice that, depending on the chosen penalty, more than one complexity parameter could be involved in the definition of $p(\boldsymbol{B} ; \lambda)$ (see Section \ref{sec:penalties} for further details).

A general-purpose algorithm for maximizing \eqref{eq:pen_log_lik} can be devised, as described in the next subsection.

\subsection{Model estimation}\label{sec:em_algo}
Direct maximization of \eqref{eq:pen_log_lik} is unfeasible, as the terms $\vect \left(\boldsymbol{\Lambda}_j\right)$, $j=1,\ldots,J$, are unknown. We therefore devise an EM algorithm \citep{Dempster1977} in which the E-step computes the conditional expectations for the unobserved quantities,  
 while a \textit{complete penalized log-likelihood} is maximized  in the M-step.
\subsubsection{E-step}
The E-step requires the computation of $\mathbb{E}(\vect \left(\boldsymbol{\Lambda}_j\right)|\Y_j;\boldsymbol{\theta})$ and $\mathbb{E}(\vect \left(\boldsymbol{\Lambda}_j\right)\vect \left(\boldsymbol{\Lambda}_j\right)^{'}|\Y_j;\boldsymbol{\theta})$. This is achieved by noticing that the conditional density $p(\vect \left(\boldsymbol{\Lambda}_j\right)| \Y_j;\boldsymbol{\theta})$ is Normal. Updating formulae for the quantities of interest are thus derived as follows:

\begin{equation} \label{eq:e_step_var}
\hat{\GAMMA}_j=\mathbb{V}(\vect \left(\boldsymbol{\Lambda}_i\right)|\Y_j;\boldsymbol{\theta})=\left[\left( \boldsymbol{I}_{r} \otimes  \boldsymbol{Z}_{j} \right)^{'} \left( \SIGMA \otimes \boldsymbol{I}_{n_j} \right)^{-1}\left( \boldsymbol{I}_{r} \otimes  \boldsymbol{Z}_{j} \right)+\PSI^{-1}\right]^{-1},
\end{equation}
\begin{equation} \label{eq:e_step_mean}
\widehat{\vect \left(\boldsymbol{\Lambda}_j\right)}=\mathbb{E}(\vect \left(\boldsymbol{\Lambda}_j \right)|\Y_j;\boldsymbol{\theta})=\hat{\GAMMA}_j \left( \boldsymbol{I}_{r} \otimes  \boldsymbol{Z}_{j} \right)^{'} \left( \SIGMA \otimes \boldsymbol{I}_{n_j} \right)^{-1}\left(\vect \left(\boldsymbol{Y}_j \right)-(\boldsymbol{I}_{r} \otimes \boldsymbol{X}_{j}) \vect \left( \boldsymbol{B} \right)\right).
\end{equation}
Consequently, the second moment $\hat{\R}_j=\mathbb{E}(\vect \left(\boldsymbol{\Lambda}_j\right)\vect \left(\boldsymbol{\Lambda}_j\right)^{'}|\Y_j;\boldsymbol{\theta})$ reads:
\begin{equation} \label{eq:e_step_second_moment}
\hat{\R}_j=\hat{\GAMMA}_j+\widehat{\vect \left(\boldsymbol{\Lambda}_j\right)}\widehat{\vect \left(\boldsymbol{\Lambda}_j\right)}^{'}.
\end{equation}
At the $t$-th iteration of the EM algorithm, the E-step requires the computation of \eqref{eq:e_step_var}-\eqref{eq:e_step_second_moment} conditioning on the parameter values estimated at iteration $t-1$. Notice that we can directly define the conditional density of $\Y_j|\boldsymbol{\Lambda}_j$ by means of the matrix normal distribution
\begin{equation} \label{eq:matr_cond_distribution}
\Y_j|\boldsymbol{\Lambda}_j \sim m\mathcal{N}\left( \boldsymbol{X}_{j} \boldsymbol{B}+\boldsymbol{Z}_{j} \boldsymbol{\Lambda}_{j}, \boldsymbol{I}_{n_j},\SIGMA\right),
\end{equation}
where $\boldsymbol{X}_{j} \boldsymbol{B}+\boldsymbol{Z}_{j} \boldsymbol{\Lambda}_{j}$ is the $n_j \times r$ mean matrix, and $\boldsymbol{I}_{n_j}$, $\SIGMA$ respectively identify the row and column covariance matrices \citep{Dawid1981}.  Such a representation will be useful in specifying the update for $ \boldsymbol{B}$ in the devised M-step: details are provided in the next subsection.

\subsubsection{M-step} \label{sec:M_step}
In the M-step we maximize the \textit{complete penalized log-likelihood}:
\begin{multline}\label{eq:complete_pen_ll}
\ell_{C\:pen}(\boldsymbol{\theta})=\sum_{j=1}^J \log(p(\vect\left(\Y_j\right)|\vect \left(\boldsymbol{\Lambda}_j\right);\boldsymbol{B},\SIGMA))+\log(p(\vect \left(\boldsymbol{\Lambda}_j\right);\PSI))-p(\boldsymbol{B} ; \lambda)=\\
=\sum_{j=1}^J-\frac{n_j}{2}\log(2\pi)-\frac{1}{2}\log| \SIGMA \otimes \boldsymbol{I}_{n_j} |-\frac{1}{2}\mathbb{E}(\e_j^{'}\left(  \SIGMA \otimes \boldsymbol{I}_{n_j} \right)^{-1}\e_j|\Y_j,\boldsymbol{\theta})+\\ 
-\frac{n_j}{2} \log(2 \pi)-\frac{1}{2} \log|\PSI|-\frac{1}{2} \mathbb{E}\left(\vect \left(\boldsymbol{\Lambda}_j\right)^{'} \PSI^{-1} \vect \left(\boldsymbol{\Lambda}_j\right)|\Y_j,\boldsymbol{\theta}\right)-p(\boldsymbol{B} ; \lambda),
\end{multline}
where $\e_j=\vect\left(\Y_j\right)-(\boldsymbol{I}_{r} \otimes \boldsymbol{X}_{j}) \vect \left( \boldsymbol{B} \right)-(\boldsymbol{I}_{r} \otimes \boldsymbol{Z}_{j}) \vect \left( \boldsymbol{\Lambda}_j \right)$ and the maximization is performed with respect to $\boldsymbol{\theta}=\{\boldsymbol{B}, \SIGMA,\PSI\}$. 

The updating formula for $\boldsymbol{B}$ clearly depends on the considered $p(\boldsymbol{B} ; \lambda)$ penalty.  All the same, it is convenient to work with the matrix-variate representation defined in  \eqref{eq:matr_cond_distribution}. In so doing, the objective function to be maximized wrt $\boldsymbol{B}$ reads:

\begin{equation} \label{eq:m_step_B}
Q_{\boldsymbol{B}}(\boldsymbol{B})=-\frac{1}{2}\sum_{j=1}^J \operatorname{tr}\left(\SIGMA^{-1} \left(\tilde{\boldsymbol{Y}}_{j}-\boldsymbol{X}_{j} \boldsymbol{B}\right)^{'}\left(\tilde{\boldsymbol{Y}}_{j}-\boldsymbol{X}_{j} \boldsymbol{B}\right) \right)-p(\boldsymbol{B} ; \lambda),
\end{equation}
where $\tilde{\boldsymbol{Y}}_{j}=\Y_j-\boldsymbol{Z}_{j} \hat{\boldsymbol{\Lambda}}_{j}$. $\hat{\boldsymbol{\Lambda}}_{j}$ is recovered by applying the inverse of the vectorization operator to $\widehat{\vect \left(\boldsymbol{\Lambda}_j\right)}$, previously computed in the E-step. Simply put, the $\widehat{\vect \left(\boldsymbol{\Lambda}_j\right)}$ vector of length $qr$ is rearranged in a $q \times r$ matrix, obtaining $\hat{\boldsymbol{\Lambda}}_{j}$. Start by noticing that, when no penalty is considered, maximization of \eqref{eq:m_step_B} agrees with the generalized least squares (GLS) estimator assuming $\SIGMA$ and $\PSI$  known \citep{Shah1997}. By exploiting properties of the trace operator, we can rewrite \eqref{eq:m_step_B} defining the following minimization problem:
\begin{equation} \label{eq:m_step_B_F_norm}
\operatorname{minimize}_{\boldsymbol{B} \in \mathbb{R}^{p\times r}} \frac{1}{2}\sum_{j=1}^J \left| \left| \SIGMA^{-1/2} \left(\tilde{\boldsymbol{Y}}_{j}-\boldsymbol{X}_{j} \boldsymbol{B}\right)^{'}\right|\right|^2_F +p(\boldsymbol{B} ; \lambda),
\end{equation}
where $||\cdot||^2_F$ denotes the squared Frobenius norm and $ \SIGMA^{-1/2}$ is the symmetric positive definite square root of $ \SIGMA^{-1}$, such that $\SIGMA^{-1}=\SIGMA^{-1/2}\SIGMA^{-1/2}$. The representation in \eqref{eq:m_step_B_F_norm} allows to employ standard routines for multivariate penalized fixed-effects models for estimating $\boldsymbol{B}$. In details, for solving \eqref{eq:m_step_B_F_norm} a two-step updating scheme is devised. Firstly, we compute
\begin{equation} \label{eq:M_step_fixed}
\tilde{\boldsymbol{B}}=\argmin_{\boldsymbol{B}} \frac{1}{2}\sum_{j=1}^J \left| \left| \SIGMA^{-1/2} \tilde{\boldsymbol{Y}}_{j}-\boldsymbol{X}_{j} \boldsymbol{B}\right|\right|^2_F +p(\boldsymbol{B} ; \lambda),
\end{equation}
that is a fixed-effects penalized regression problem in which the response variable is $\SIGMA^{-1/2} \tilde{\boldsymbol{Y}}_{j}$, $j=1,\ldots,J$; $\tilde{\boldsymbol{B}}$ is thus easily retrieved via fixed-effects routines for penalized estimation. Secondly, the solution to \eqref{eq:m_step_B_F_norm} is obtained  post multiplying $\tilde{\boldsymbol{B}}$ by $\SIGMA^{1/2}$. Therefore, at each iteration of the EM-algorithm, we firstly compute $\tilde{\boldsymbol{B}}$ and then we set:
\begin{equation}
\hat{\boldsymbol{B}}=\tilde{\boldsymbol{B}}\SIGMA^{1/2},
\end{equation}
where $\hat{\boldsymbol{B}}$ maximizes \eqref{eq:m_step_B}. This procedure stems from the rationale outlined in \cite{Rohart2014}, where, contrarily to their original solution, in our context the updating steps are made more complex by the multidimensional nature of $\Y$. The devised updating scheme allows to easily incorporate any $p(\boldsymbol{B} ; \lambda)$ that has been previously defined for the fixed-effects framework, and whose estimating routines are available. A list of possible penalties is proposed in Section \ref{sec:penalties}.

Updating formulae for the covariance matrices $\PSI$ and $\SIGMA$ agree with those of the unpenalized framework, namely
\begin{equation}
\hat{\PSI}=\frac{1}{J}\sum_{j=1}^J\hat{\R}_{j},
\end{equation}
and for the $(h, k)$-th element of matrix $\SIGMA$
\begin{equation} \label{eq:update_sigma}
\hat{\SIGMA}_{(h,k)}=\frac{1}{N}\sum_{j=1}^J \left[ \mathbb{E}\left(\boldsymbol{E}_{jh}|\Y_j\right)^{'}  \mathbb{E}\left(\boldsymbol{E}_{jk}|\Y_j\right)\right]+\operatorname{tr}\left[ \operatorname{cov}(\boldsymbol{E}_{jh},\boldsymbol{E}_{jk}|\Y_j)\right], \quad h,k= 1,\ldots,r,
\end{equation}
where $\boldsymbol{E}_{jh}$ denotes the $h$-th column of matrix $\boldsymbol{E}_{j}=\Y_j-\boldsymbol{Z}_{j} \hat{\boldsymbol{\Lambda}}_{j}-\boldsymbol{X}_{j} \boldsymbol{B}$, $h=1,\ldots,r$. 
\subsection{Definition of $p(\boldsymbol{B} ; \lambda)$} \label{sec:penalties}
The EM algorithm devised in the previous section defines a general-purpose optimization strategy for penalized mixed-effects multitask learning. While any penalty type can in principle be defined, three notable examples, commonly used in this context, are the elastic net penalty \citep{Zou2005}, the multivariate group-lasso penalty \citep{Obozinski2011} and the netReg routines for Network-regularized linear models \citep{Dirmeier2018}. Each of them is briefly described in the next subsections.
\subsubsection{Elastic-net penalty} \label{sec:elnet_penalty}
The first penalty type we consider is the renowned convex combination of lasso and ridge regularizers, whose magnitude of the former over the latter is controlled by the mixing parameter $\alpha$, $0 \leq \alpha \leq 1$. In details, the penalty expression reads:
\begin{equation} \label{eq:elnet_penalty}
p(\boldsymbol{B} ; \lambda, \alpha)=\lambda\left[(1-\alpha)\sum_{c=1}^r\sum_{l=2}^p b_{lc}^2 +\alpha \sum_{c=1}^r\sum_{l=2}^p|b_{lc}|\right],
\end{equation}
where $b_{lc}$ denotes the element in the $l$-th row and $c$-th column of matrix $\boldsymbol{B}$. Notice that the first row of $\boldsymbol{B}$ contains the $r$ intercepts and it is thus not penalized. Algorithmically, penalty \eqref{eq:elnet_penalty} can be enforced employing standard and widely available routines for univariate penalized estimation, like the \texttt{glmnet} software \citep{Tay2021}. The only computational detail that shall be examined is how to prevent the default shrinkage of the $r$ intercepts: the \texttt{penalty.factor} argument of the  \texttt{glmnet} function effectively serves the purpose. \re{The latter can also be employed in our framework to force coefficients that need not be penalized to enter the model specification.}

\subsubsection{Multivariate group-lasso penalty} \label{sec:group_lasso}
This type of penalty imposes a group structure on the coefficients, forcing the same subset of predictors to be preserved across all $r$ components of the response matrix. This feature is particularly desirable when building multivariate DNAm biomarkers, since it automatically identifies the CpG sites that are \textit{jointly} related to the considered risk factors. Such a penalty is defined as follows:

\begin{equation} \label{eq:group_lasso_penalty}
p(\boldsymbol{B} ; \lambda, \alpha)=\lambda\left[(1-\alpha)\sum_{c=1}^r\sum_{l=2}^p b_{lc}^2 +\alpha \sum_{l=2}^p||\boldsymbol{b}_{l.}||_2\right],
\end{equation}
where $\boldsymbol{b}_{l.}$ identifies the $l$-th row of the matrix $\boldsymbol{B}$, such that each $\boldsymbol{b}_{l.}$, $l=2,\ldots,p$ is an $r$-dimensional vector. Likewise Section \ref{sec:elnet_penalty}, summations over rows in \eqref{eq:group_lasso_penalty} start at $2$ since we do not penalize the vector of intercepts. This penalty  behaves like the lasso, but on the whole group of predictors for each of the $r$ variables: they are either all zero, or else none are zero, but are shrunk by an amount depending on $\lambda$. Similarly to \eqref{eq:elnet_penalty}, the mixing parameter $\alpha$ controls the weight associated to 
ridge and group-lasso regularizers.
The \texttt{glmnet} software, with \texttt{family = "mgaussian"} is again at our disposal for efficiently incorporating \eqref{eq:group_lasso_penalty}  in the framework outlined in the present paper. 

\subsubsection{Network-Regularized penalty} \label{sec:netreg}
The last penalty we consider allows for the inclusion of biological graph-prior knowledge in the estimation by  accounting for the contribution of two non-negative adjacency matrices $\boldsymbol{G}_X \in \mathbb{R}_{+}^{(p-1)\times (p-1)}$ and $\boldsymbol{G}_Y \in \mathbb{R}_{+}^{r \times r}$, respectively related to $\X$ and $\Y$. In this case, $p(\boldsymbol{B} ; \lambda)$ assumes the following functional form:

\begin{equation} \label{eq:netreg_penalty}
p(\boldsymbol{B} ; \lambda,\lambda_X,\lambda_Y)=\lambda ||\boldsymbol{B}_0||_1+
\lambda_{X} \tr \left(\boldsymbol{B}_0^{'}(\boldsymbol{D}_{G_X}-\boldsymbol{G}_X)\boldsymbol{B}_0 \right)+
\lambda_{Y}\tr \left(\boldsymbol{B}_0(\boldsymbol{D}_{G_Y}-\boldsymbol{G}_Y)\boldsymbol{B}_0^{'} \right),
\end{equation}
where $\boldsymbol{B}_0$ is the $(p-1) \times r$ matrix of coefficients without the intercepts and $\boldsymbol{D}_{G_X}$, $\boldsymbol{D}_{G_Y}$ indicate the degree matrices of $\boldsymbol{G}_X$ and $\boldsymbol{G}_Y$, respectively \citep{Chung1997}. $\boldsymbol{G}_X$ and $\boldsymbol{G}_Y$ encode a biological similarity, forcing rows and columns of $\boldsymbol{B}_0$ to be similar. The \texttt{netReg R} package  \citep{Dirmeier2018} provides a convenient implementation of \eqref{eq:netreg_penalty}.

\subsubsection{On the choice of $p(\boldsymbol{B} ; \lambda)$} \label{sec:comparison_pen}
\re{Leaving the flexibility attained by the methodology proposed in Section \ref{sec:mod_def} aside; in practice, a functional form for $p(\boldsymbol{B} ; \lambda)$ must be chosen when performing the analysis. We hereafter highlight pros and cons of the proposed approaches with respect to a mixed-effects multitask learning setting.}

\re{The elastic-net penalty in \eqref{eq:elnet_penalty} does not take into account the multivariate nature of the problem in \eqref{eq:pen_log_lik}, as the shrinkage is applied directly to $\vect(\boldsymbol{B})$. This behavior allows for capturing a wide variety of sparsity patterns that may be present in $\boldsymbol{B}$, but does not impose any specific structure that could be desirable in a multivariate context. Differently, the multivariate group-lasso of Section \ref{sec:group_lasso} defines a shrinkage term that forces the same subset of predictors to be preserved across all $r$ components of the response $\boldsymbol{Y}$. This can be seen as the generalization of the variable selection problem to the multivariate response setting, which is also known as \textit{support union problem} or \textit{row selection problem} in the literature \citep{Obozinski2011}. Lastly, the network-Regularized penalty in \ref{sec:netreg} is particularly useful when the interaction among features and/or responses is, at least partially, known, such that it can be profited from within the learning mechanism \citep{Cheng2014}.}

\re{In relation to the DNAm surrogate creation task motivating our methodological proposal, the multivariate group-lasso is definitely the most appropriate penalty, as it not only showcases better prediction performances but it is also supported by biological reasons: a thorough analysis for the EPIC dataset is reported in Section \ref{sec:application}.
}
\subsection{Further aspects} \label{sec:further_aspect}
Hereafter, we discuss some practical considerations related to the presented methodology.
\begin{itemize}
\item \textbf{Initialization:} we start the algorithm with an M-step, setting $\hat{\theta}^{(0)}=\{\hat{\boldsymbol{B}}^{(0)}, \hat{\SIGMA}^{(0)},\hat{\PSI}^{(0)}\}$. In details, both $\hat{\SIGMA}^{(0)}$ and $\hat{\PSI}^{(0)}$ are initialized with identity matrices of dimension $r \times r$ and $qr \times qr$ respectively, while  $\hat{\boldsymbol{B}}^{(0)}$ is estimated from a penalized linear model (without the random-effects) employing the chosen penalty function with the associated hyper-parameters.
\item \textbf{Convergence:} the EM algorithm is considered to have converged once the relative difference in the objective function for two subsequent iterations is smaller than $\varepsilon$, for a given $\varepsilon>0$:
\[\frac{|\ell_{pen}(\hat{\boldsymbol{\theta}}^{(t+1)})-\ell_{pen}(\hat{\boldsymbol{\theta}}^{(t)})|}{|\ell_{pen}(\hat{\boldsymbol{\theta}}^{(t)})|}<\varepsilon,\]
where $\hat{\boldsymbol{\theta}}^{(t)}=\{\hat{\boldsymbol{B}}^{(t)}, \hat{\SIGMA}^{(t)},\hat{\PSI}^{(t)}\}$ is the set of estimated values at the end of the $t$-th iteration. In our analyses, $\varepsilon$ is set equal to $10^{-6}$. The procedure described in Section \ref{sec:em_algo} falls within the class of Expectation Conditional Maximization (ECM) algorithms, whose convergence properties have been proved in \cite{Meng1993} and in Section 5.2.3 of \cite{McLachlan2008}.

\item \textbf{Model selection:} a standard 10-fold cross validation (CV) strategy is employed for selecting the tuning factors. 
Alternatively, as suggested in \cite{Rohart2014}, one could employ a modified version of the Bayesian Information Criterion \citep[BIC,][]{Schwarz1978}:
\begin{eqnarray} \label{eq:modified_BIC}
B I C=2 \ell(\hat{\boldsymbol{\theta}})-d_{0}\log (N), 
\end{eqnarray}
where $\ell(\hat{\boldsymbol{\theta}})$ is the log-likelihood evaluated at $\hat{\boldsymbol{\theta}}$, obtained maximizing \eqref{eq:pen_log_lik}, and $d_{0}$ is the number of non-zero parameters resulting from the penalized estimation. Another option would be to rely on an interval search algorithm, like the efficient parameter selection via global optimization  \citep{Frohlich2005}: an implementation is available in the \texttt{c060 R} package \citep{Sill2014a}. 
\item \textbf{Scalability:} the devised methodology provides a framework for incorporating any penalty in a high-dimensional mixed-effects multitask learning framework. To this extent, the data dimensionality our procedure can cope with, \re{as well as the overall computing time,} very much depends on the scalability and \re{efficiency} associated to the chosen shrinkage term. Typically nevertheless, penalized likelihood approaches fail to be directly applied to ultrahigh-dimensional problems \citep{Fan2009}, and pre-processing procedures such as variable screening are thus required prior to modeling. The epigenetic application that motivated our work naturally called for an EWAS pre-screening strategy (see Section \ref{sec:EPIC_data}), but clearly other dimensionality reduction techniques could be considered when dealing with massive datasets. The interested reader is referred to \cite{Jordan2013} for a thought-provoking investigation on the topic.
\item \textbf{Implementation:} routines for fitting the penalized mixed-effects multitask learning method have been implemented in \texttt{R} \citep{RCoreTeam}, and the source code is freely available in the Supplementary Material and at \texttt{https://github.com/AndreaCappozzo/emlmm} in the form of an \texttt{R} package. The three penalties described in Section \ref{sec:penalties} are included in the software, and can be selected via the \texttt{penalty\_type} argument of the \texttt{ecm\_mlmm\_penalized} function. As described in Section \ref{sec:penalties}, the M-step heavily relies on previously developed fast and stable subroutines, while the E-step and the objective function evaluation have been implemented in \texttt{c++} to reduce the overall computing time. 
\item \textbf{Response-specific random-effects:} model in \eqref{eq:r_dim_lmm} assumes that each and every response requires a random-effects component. Whilst in principle reasonable, it may happen in specific applications that only a subset of the $r$ characteristics in $\Y$ enjoys group-dependent heterogeneity. The occurrence of such a scenario can be unveiled by looking at the $r$ diagonal elements of dimension $q$ in $\hat{\boldsymbol{\Psi}}$: a response may be considered group-independent when the magnitude of the associated elements in $\text{diag}(\hat{\boldsymbol{\Psi}})$ is significantly lower than the remaining ones. Doing this way,  the impact a given random-effect has on the different characteristics is retrieved as a by-product of the modeling procedure.

\end{itemize}

\section{Simulation study} \label{sec:sim_study}
In this section, we evaluate the model introduced in Section \ref{sec:methodology} on synthetic data. The aim of the analyses reported hereafter is twofold. On the one hand, we would like to validate the predictive power of the proposed procedure against its fixed-effects counterpart when the random-effects vary across dimensions in the multivariate response. On the other hand, we assess the estimated model parameters and the recovery of the underlying sparsity structure for different values of the shrinkage factor $\lambda$.
\subsection{Experimental setup}
We generate $N=600$ data points according to model \eqref{eq:r_dim_lmm} with the following parameters:
\[
\PSI= \begin{bmatrix}{}
  50.00 & -1.59 & -0.60 & -0.22 & 2.38 \\ 
  -1.59 & 40.00 & -0.96 & -0.91 & 0.37 \\ 
  -0.60 & -0.96 & 30.00 & -0.43 & 0.50 \\ 
  -0.22 & -0.91 & -0.43 & 20.00 & 0.80 \\ 
  2.38 & 0.37 & 0.50 & 0.80 & 0.16 \\ 
  \end{bmatrix}, \quad \SIGMA=\begin{bmatrix}{}
  2.16 & 0.09 & -0.80 & 0.91 & -0.26 \\ 
  0.09 & 2.16 & -0.33 & 0.55 & -0.10 \\ 
  -0.80 & -0.33 & 2.16 & -0.03 & -0.13 \\ 
  0.91 & 0.55 & -0.03 & 2.16 & 0.02 \\ 
  -0.26 & -0.10 & -0.13 & 0.02 & 2.16 \\ 
  \end{bmatrix},\]
implying that $r=5$ and $q=1$. Notice that $\PSI$ is purposely constructed for the random-effects to differently affect the five dimensional response: while the first component showcases high variance (first entry in the main diagonal) the last one is very small and close to $0$. \re{Further, the error variances (diagonal elements of $\SIGMA$) are held constant across dimensions to better highlight the impact the variability of the random-effects has on the models performance.} The data generating process assumes ten equally-sized subpopulations, resulting in $J=10$. The matrix of fixed-effects $\boldsymbol{B}$ is of dimension $10001 \times 5$, with distinct sparsity pattern according to three scenarios:
\begin{itemize}
\item \textit{$\boldsymbol{B}$ row-wise sparse}: $\boldsymbol{B}$ has entries equal to $0.5$ for the first $100$ rows, while all the other entries are equal to $0$,
\item \textit{$\boldsymbol{B}$ sparse at random}: $\boldsymbol{B}$ is equal to $0.5$ for approximately $70\%$ of its entries, while all the others are equal to $0$,
\item \re{\textit{$\boldsymbol{B}$ with dependence structure}: $\boldsymbol{B}$ has entries whose magnitude agrees with the correlation structure between the covariates, inducing coefficients to be similar when the absolute correlation between two predictors is high.}
\end{itemize}
 Lastly, $\boldsymbol{Z}_{j}$ is an all-one column vector $\forall j=1,\ldots,10$, while $\boldsymbol{X}_{j}$ has the first column equal to $1$, meaning that the intercept is included in $\boldsymbol{X}_{j}$ in our model specification. \re{The remaining $10000$ dimensions are generated according to a normal random vector with independent marginals for the \textit{$\boldsymbol{B}$ row-wise sparse} and \textit{$\boldsymbol{B}$ sparse at random} scenarios, while the \texttt{cormat\_from\_triangle} function from the \texttt{faux} package \citep{DeBruine2021} has been used to simulate correlated predictors in the \textit{$\boldsymbol{B}$ with dependence structure} experiment.}

\begin{figure}
\centering
    \includegraphics[width=\linewidth]{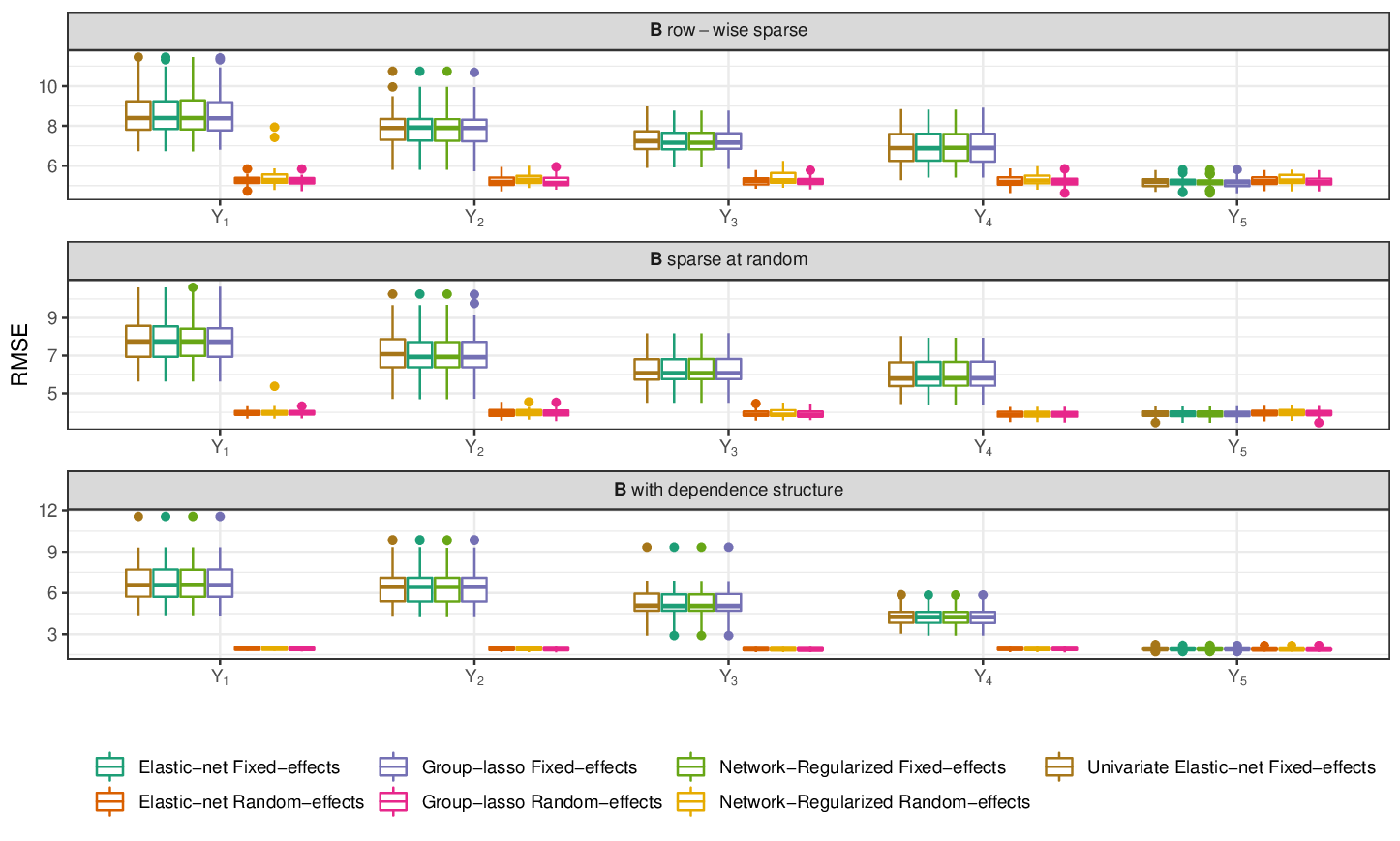}
    \caption{Boxplots of the Root Mean Squared Error (RMSE) for $MC = 100$ repetitions of the simulated experiment. RMSE is computed on $200$ test points for different methods and three scenarios varying sparsity pattern for $\boldsymbol{B}$.}
  \label{fig:rmse_sim_study}
\end{figure}

Taking a cue from the Monte Carlo simulations of \cite{Li2010}, for each replication of our experiment the learning framework is structured as follows: we equally divide the $N=600$ units in a training set, an independent validation set and an independent test set, retrieving a sample size of $200$ for each.  Notice that, as to mimic the process of DNAm surrogates creation, the total number of variables ($p=10001$) is much larger than the sample size. Seven different models, varying $\lambda$ within a grid, are fitted on the training data:
\begin{itemize}
\item \re{\textit{Univariate Elastic-net Fixed-effects:} univariate elastic-net regression, obtained fitting independent models to each dimension of the multivariate response,}
\item \textit{Elastic-net Fixed-effects:} a penalized multitask learning model with elastic-net regularization. The considered penalty is described in Section \ref{sec:elnet_penalty},
\item \textit{Group-lasso Fixed-effects:} a penalized multitask learning model with multivariate group-lasso regularization. The considered penalty is described in Section \ref{sec:group_lasso},
\item \textit{Network-Regularized Fixed-effects:} graph-regularized multitask learning model with edge-based regularization. The considered penalty is described in Section \ref{sec:netreg},
\item \textit{Elastic-net Random-effects:} the penalized MLMM methodology introduced in the paper with elastic-net regularization (Section \ref{sec:elnet_penalty}),
\item \textit{Group-lasso Random-effects:} the penalized MLMM methodology introduced in the paper, with group-lasso regularization (Section \ref{sec:group_lasso}),
\item \textit{Network-Regularized Random-effects:} the penalized MLMM methodology introduced in the paper, with edge-based regularization (Section \ref{sec:netreg}).
\end{itemize}
\re{Such an extensive comparison can be regarded as performing a within-scenario ablation study, in which we start from a complex method and we subsequently remove the random-effects component and, finally, the borrow strength property of multivariate regression, to be left with Univariate Elastic-net Fixed-effects models. In this way, we investigate the contribution of our proposal to the overall system.} The mixing parameter $\alpha$ was set equal to $0.5$ for methods with elastic-net and group-lasso regularizers, while for the Network-Regularized penalty we employ $5$-fold CV to tune $\lambda_{X}$ and $\lambda_{Y}$ on the training set. For the latter penalty, the adjacency matrices $\boldsymbol{G}_X$ and $\boldsymbol{G}_Y$ are computed via a thresholding procedure on the correlation matrices of $\boldsymbol{X}$ and $\boldsymbol{Y}$, respectively, with a threshold equal to $0.1$ \citep{Langfelder2008}. Subsequently, the validation dataset is used to select the best shrinkage parameter $\lambda$ minimizing the RMSE for every model. The predictive performance is then evaluated on the test set. \re{Lastly, to assess out of groups prediction, models are further validated on $100$ external samples, generated according to \eqref{eq:r_dim_lmm}, coming from five extra subpopulations  not observed in the training
set.}
The devised simulated experiment is replicated $MC=100$ times: results are reported in the next subsection.
\subsection{Simulation results} 

Figure \ref{fig:rmse_sim_study} displays boxplots of the Root Mean Squared Error, computed for each component of the $5$-dimensional response on the test set. For all scenarios we observe that the component-wise predictive performance is heavily affected by the magnitude of the related diagonal entry in the $\boldsymbol{\PSI}$ matrix. When the grouping effect is negligible (fifth dimension $Y_5$), all methods showcase comparable predictive performance under both scenarios. Contrarily, the RMSE deteriorates for fixed-effects models in those response components for which the grouping impact is more relevant. The same does not happen for the mixed-effects counterparts, as the random intercept effectively captures baseline differences across groups. Interestingly, the penalty type does not seem to influence the RMSE metric, with our proposal displaying excellent results irrespective of the chosen shrinkage functional for all scenarios. On the other hand, when it comes to perform out of groups prediction the gain achieved by including random-effects decreases and the outcome of models with fixed and mixed-effects are fairly similar. In details, for the latter class of methods the unconditional (population level) intercepts are employed when making predictions for unobserved groups. Notwithstanding, we recognize that results are no worse than those obtained with fixed-effects procedures, corroborating the generalizability of our proposals in external cohorts.

\begin{figure}
\centering
    \includegraphics[scale=.7]{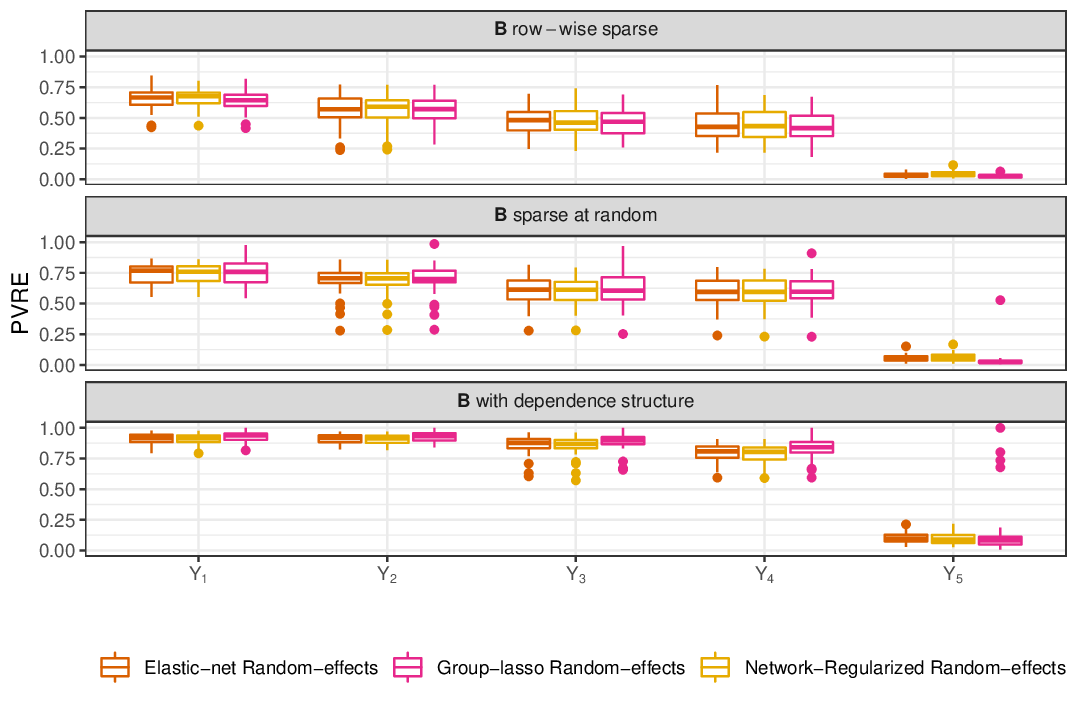}
    \caption{Boxplots of the Percentage of Variation due to Random Effects (PVRE) for $MC = 100$ repetitions of the simulated experiment.}
  \label{fig:PVRE_sim_study}
\end{figure}

Figure \ref{fig:PVRE_sim_study} displays the analogue of the Percentage of Variation due to Random Effects (PVRE) metric,
computed taking the ratio between the diagonal elements of $\hat{\boldsymbol{\PSI}}$ and the sum of the diagonals of $\hat{\boldsymbol{\PSI}}$ and $\hat{\boldsymbol{\Sigma}}$. From the plot, it clearly emerges how the grouping impact differently affects the variability in the five components of the response. 

\begin{figure}
\centering
    \includegraphics[width=\linewidth]{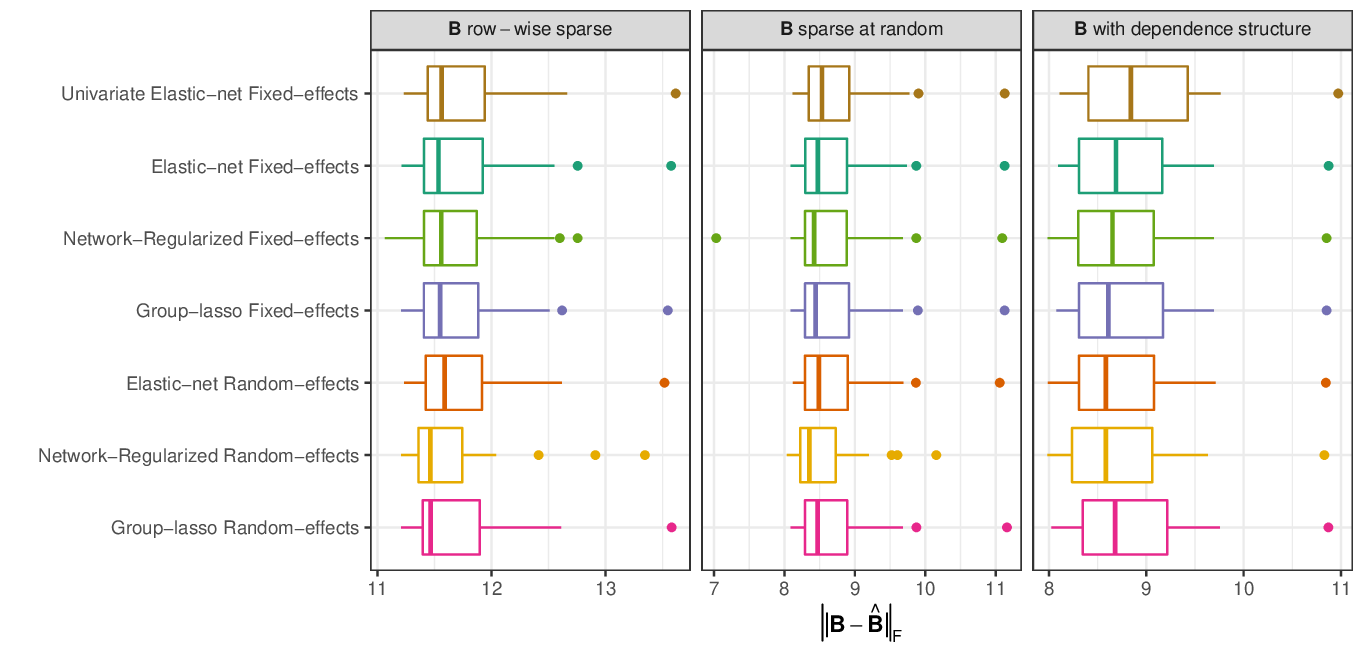}
    \caption{Boxplots of the Frobenius distance between true and estimated matrices of fixed-effects $\boldsymbol{B}$ for $MC = 100$ repetitions of the simulated experiment.}
  \label{fig:beta_frobenius_sim_study}
\end{figure}

\re{Figure \ref{fig:beta_frobenius_sim_study} reports boxplots of the Frobenius distance between true and estimated matrices of fixed-effects $\boldsymbol{B}$. When looking at $||\boldsymbol{B}-\hat{\boldsymbol{B}}||_F$ under the three scenarios we observe some interesting facts.  First off, it is immediately noticed that the \textit{Univariate Elastic-net Fixed-effects} model showcases the poorest performance, in particular for the \textit{$\boldsymbol{B}$ with dependence structure} experiment. This is due to the fact that the different components of the response vector are related in our simulated specification, and therefore fitting separate regression models results in a loss of quality for the estimator. Secondly, we observe that for the \textit{$\boldsymbol{B}$ row-wise sparse} scenario the \textit{Group-lasso Random-effects} model is the best performing one among all the competitors, displaying the lowest median distance to $\boldsymbol{B}$. This may be expected, as such method is precisely constructed to identify a matrix of fixed-effects with row-wise sparsity patterns. Furthermore, notice that the performance of the \textit{Group-lasso Random-effects} is slightly better than its Fixed-effects counterpart. For the remaining scenarios the superiority of the mixed-effects procedures is not so apparent, and both fixed and random-effects models demonstrate a comparable performance. An only modest gain is showcased by the \textit{Network-Regularized Random-effects} method, for which the inclusion of the adjacency matrices $\boldsymbol{G}_X$ and $\boldsymbol{G}_Y$ in the penalty specification helps in better recovering the $\boldsymbol{B}$ structure.}

\begin{figure}
\centering
    \includegraphics[width=\linewidth]{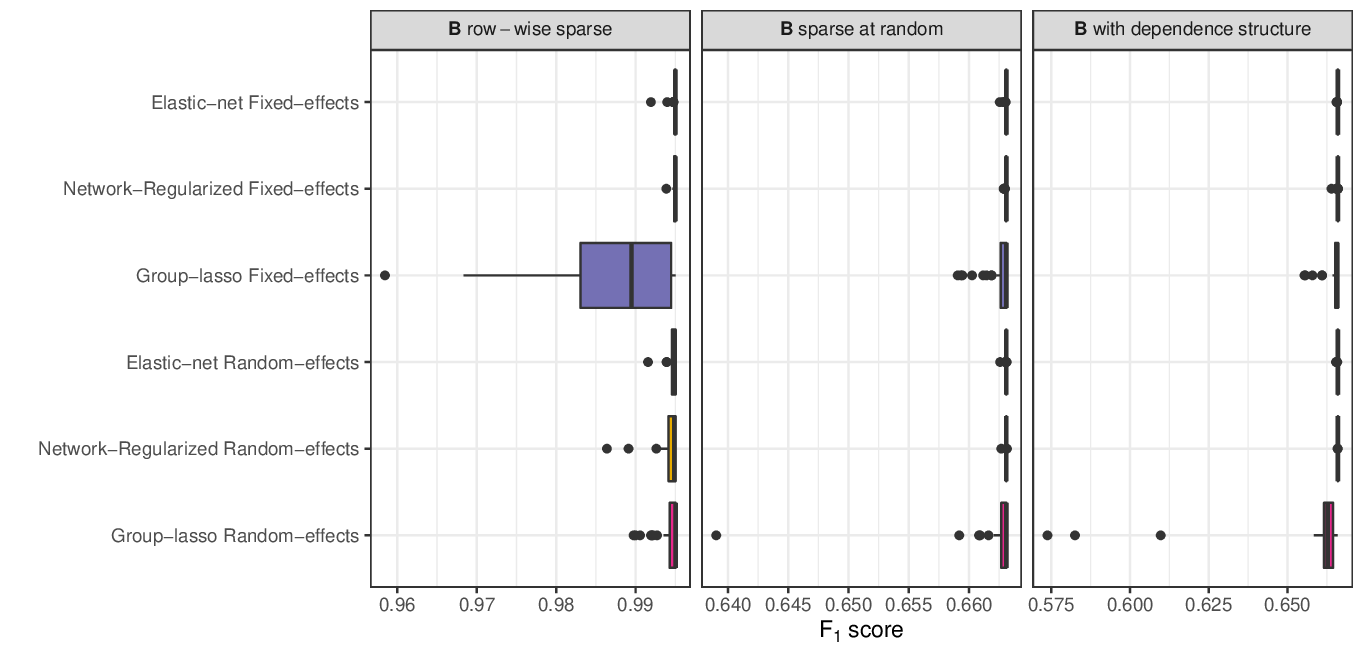}
    \caption{Boxplots of the $F_1$ score for $MC = 100$ repetitions of the simulated experiment for different methods and three scenarios varying sparsity pattern for $\boldsymbol{B}$.}
  \label{fig:F1_sim_study}
\end{figure}

\re{We now look at the ability of the competing procedures in recovering the true underlying sparsity patterns in the matrix of fixed-effects $\boldsymbol{B}$ under the different scenarios. In so doing we compute, for each replication of the simulated experiment, the $F_1$ score defined as follows:
\begin{equation} \label{eq:f1_measure}
F_1=\frac{\texttt{tp}}{\texttt{tp}+0.5(\texttt{fp}+\texttt{fn})},
\end{equation}
where with $\texttt{tp}$  we denote the number of zero entries in $\boldsymbol{B}$ correctly estimated as such, while $\texttt{fp}$ and $\texttt{fn}$ represent the number of non-zero entries wrongly shrunk to $0$ and the number of zero entries not shrunk to $0$, respectively. Figure \ref{fig:F1_sim_study} displays boxplots of such metric for different methods and scenarios. We notice that \textit{$\boldsymbol{B}$ row-wise sparse} structure displays much higher $F_1$ score, irrespective of the considered method, than the other two cases. Intuitively, the former scenario is less challenging since, while all penalty types can potentially accommodate a row-wise sparse $\boldsymbol{B}$,  group-lasso regularizers only force entire rows of $\boldsymbol{B}$ to be shrunk to $0$. We further observe that the $F_1$ score is higher for the \textit{Group-lasso Random-effects} than for its fixed-effects counterpart, highlighting that a penalized mixed-effects modeling strategy, in presence of grouped data and a row-wise sparse $\boldsymbol{B}$, not only increases the predictive accuracy but also improves the recovery of the sparsity pattern in the fixed-effects matrix. The same does not happen in the remaining two scenarios, for which all methods display a comparable empirical distribution of the $F_1$ metric across simulations. The same behavior was already observed in high-dimensional linear mixed-effects modeling for univariate responses \citep{Schelldorfer2011}.
}

\re{As a last worthy note we acknowledge that, as rightly underlined by an anonymous reviewer, the present simulation study does not consider any violation in the distributional assumptions of the involved quantities. In this regard, we replicated the experiment using both a Multivariate skew-normal and Multivariate skew-t distributions \citep{azzalini2013skew} as generative models for the error term, but we did not report the results in the paper since no dramatic changes were observed in model performances. While clearly more extreme scenarios could be considered, results in the literature have previously validated the robustness of linear mixed-effects models to violations of distributional assumptions  \citep{McCulloch2011a}.} For further details about the simulation study, the Supplementary Material \citep{Cappozzo2023_supplement} provides additional figures and a note on the overall computing times.

\re{All in all, the good performances displayed by our proposal, particularly when coupled with a multivariate group-lasso penalty, encourage its usage in multivariate DNAm surrogates creation: promising results are reported in the next Section.}

\section{DNAm biomarkers analysis for EPIC and EXPOsOMICS datasets} \label{sec:application}
\re{The methodology described in Section \ref{sec:methodology} is employed to build a $5$-dimensional DNAm biomarker of hypertension and hyperlipidemia. As mentioned in the introduction, DNAm surrogates possess extensive advantages over their blood-measured counterparts since:
\begin{enumerate}
\item DNAm biomarkers directly account for genetic susceptibility and subject specific response to risk factors;
\item DNAm biomarkers can immediately be computed whenever DNAm values are accessible. This is particularly useful when the risk factors of interest have not been directly measured;
\item further understanding of the biomolecular mechanisms associated with complex phenotypes can be acquired through a pathway enrichment analysis \citep{Reimand2019}, allowing to identify molecular pathways overrepresented among the regressors involved in the surrogate construction (i.e., the CpG sites whose associated parameters are not shrunk to $0$).
\end{enumerate}
We subsequently assess how well the so-devised surrogates perform, for both internal and external cohorts, in reconstructing the blood measured biomarkers (Section \ref{sec:DNAm_creation}) and in predicting the clinical endpoint of interest, namely the future presence/absence of CVD events (Section \ref{sec:cvd_risk}). Lastly, we study from a biological perspective the CpG sites selection operated by the multivariate group-lasso penalty, comparing it with previous findings available in the literature (Section \ref{sec:CpG_interpretation}).}


\subsection{DNAm surrogates creation and validation} \label{sec:DNAm_creation}
\re{To construct multivariate DNAm surrogates, several penalized models are fitted to the EPIC Italy training set, varying shrinkage factors and considering both fixed and random-effects components. As mentioned in Section \ref{sec:EPIC_data}, the design matrix comprises of $p=62130$ variables. Thus, redundancies are likely to occur as the feature space is constituted by the union of CpG sites pre-screened by univariate epigenome-wide analyses. After having standardized the covariates, for each model the penalty term $\lambda$ is tuned via 10-fold CV, while the mixing parameter $\alpha$ is kept fixed and equal to $0.5$.  Results on the internal cohort are summarized in Table \ref{tab:application}, where the Root Mean Squared Error (RMSE) computed on the EPIC Italy test set, the number of active CpG sites and the overall elapsed time are reported.}
The first two rows are related to the novel penalized MLMM methodology with a random-effects design matrix that includes a $q=1$ random intercept, coupled with multivariate group-lasso (Section \ref{sec:group_lasso}) and elastic-net (Section \ref{sec:elnet_penalty}) penalties, respectively. The corresponding fixed-effects counterparts are reported in the third and fourth rows, while univariate elastic-net metrics, obtained fitting $r=5$ separate models, one for each response, are detailed in the last row of Table \ref{tab:application}. Notice that our proposal outperforms the state-of-the-art approach (univariate elastic-net) for $4$ out of $5$ dimensions of the response variable. The reason being that our method takes advantage of the borrowing information asset typical of multivariate models (the correlation between SBP and DBP is equal to $0.77$ in the training set), whilst allowing for center-wise difference to be captured by the random intercept. 
Furthermore, thanks to the multivariate group-lasso penalty, our penalized MLMM approach directly identifies the CpG sites that are jointly related to hypertension and hyperlipidemia, with a total number of features that is lower with respect to univariate elastic-nets. By taking the ratio between the diagonal elements of $\hat{\boldsymbol{\PSI}}$ and the sum of the diagonals of $\hat{\boldsymbol{\PSI}}$ and $\hat{\boldsymbol{\Sigma}}$ it is possible to compute, for each component of the response matrix $\boldsymbol{Y}$, the analogue of the Percentage of Variation due to Random Effects (PVRE) index. For the EPIC Italy dataset, the estimated PVRE amounts to $7.97\%$, $16.05\%$, $5.56\%$, $14.26\%$ and $6.01\%$ for DBP,   HDL,   LDL,   SBP and TG respectively; giving reason for the performance improvement showcased by the random-effects models.

The employment of the multivariate group-lasso penalty within a mixed-effects multitask learning framework is also supported by biological reasons. In fact, it is more likely that multiple correlated phenotypes affect (or are affected by, depending on the causal relationship between DNAm and the exposure variable) the same set of CpG sites. This mechanism is known as pleiotropic effect \citep{Tyler2014, Richard2017}. 


\begin{tiny}
\begin{table}[t]
\centering
\caption{Root Mean Squared Error (RMSE), active number of CpG sites and overall computing times for different penalized regression models, EPIC Italy test set. Bold numbers indicate lowest RMSE for each of the $r=5$ dimension of the response matrix.} 
\hspace*{-3.5cm}\begin{tabular}{llllllllll}
\hline
\multicolumn{3}{c}{Framework} &\multicolumn{5}{c}{Root Mean Squared Error} & Active \# &Elapsed time\\
Model & Penalty type & Response & DBP & HDL & LDL & SBP & TG & CpG sites &(secs)\\ 
  \hline
Random-effects & Group-lasso & Multivariate & \textbf{0.1167} & \textbf{0.2442} & \textbf{0.3236} & 0.1374 & \textbf{0.4745} & 417 & 304 \\ 
  Random-effects & Elastic-net & Multivariate & 0.1178 & 0.2480 & 0.3318 & 0.1379 & 0.4853 & 773 & 549 \\ 
  Fixed-effects & Group-lasso & Multivariate & 0.1317 & 0.2602 & 0.3321 & 0.1364 & 0.4853 & 382 & 54 \\ 
  Fixed-effects & Elastic-net & Multivariate & 0.1176 & 0.2513 & 0.3324 & 0.1362 & 0.4841 & 115 & 311 \\ 
  Fixed-effects & Elastic-net & Univariate & 0.1179 & 0.2596 & 0.3383 & \textbf{0.1359} & 0.4996 & 1712 & 115 \\ 
   \hline
\end{tabular}
\label{tab:application}
\end{table}

\end{tiny}
\re{Internal validation results obtained for the EPIC Italy test set highlight the benefits of mixed-effects modeling while constructing DNAm surrogates for data possessing a grouping-structure. The random intercept allows to account for center-wise variability that is induced by geographic genetical variation, as well as by samples collection and storage that are likely to differ across centers. Notice that in general, if not properly modeled, the unexplained heterogeneity in multi-center studies could be taken care of with dedicated batch-effect removal procedures \cite[see, e.g.,][]{Johnson2007}. Nonetheless, when developing DNAm biomarkers it is of interest to devise study-invariant surrogates, to be readily computed also for samples not belonging to the learning cohort. To this aim, we validate the performance of the models estimated on the EPIC training set in constructing surrogates for the external EXPOsOMICS cohort (see Section \ref{sec:EPIC_data}). In this context the grouping information (i.e., the center of recruitment) cannot be considered when performing predictions with mixed-effects models, and the unconditional (population-level) intercepts are thus utilized. RMSE between estimated and blood-measured biomarkers for the EXPOsOMICS validation cohort are reported in Table \ref{tab:application_EXPO}. Likewise for the EPIC Italy test set, the lowest RMSEs for all but SBP biomarker are retained employing a penalized random-intercept model with multivariate group-lasso penalty. Interestingly, the predictive outcomes obtained in the EXPOsOMICS validation dataset are comparable with the ones reported in Table \ref{tab:application}, with slightly worse performances, as it may be expected, for those dimensions of the response displaying higher PVRE indexes. All in all, the proposed approach exhibits promising results when it comes to multivariate DNAm biomarker creation, outperforming the current employed procedure in both internal and external validation cohorts.}
\begin{table}[t]
\centering
\caption{Root Mean Squared Error (RMSE) for different penalized regression models, EXPOsOMICS validation set. Bold numbers indicate lowest RMSE for each of the $r=5$ dimension of the response matrix.} 
\hspace*{-1.5cm}\begin{tabular}{lllrrrrr}
  \hline
Model & Penalty type & Response & DBP & HDL & LDL & SBP & TG \\ 
  \hline
Random-effects & Group-lasso & Multivariate & \textbf{0.1314} & \textbf{0.2384} & \textbf{0.2707} & 0.1412 & \textbf{0.4735} \\ 
  Random-effects & Elastic-net & Multivariate & 0.1335 & 0.2504 & 0.2821 & 0.1450 & 0.4890 \\ 
  Fixed-effects & Group-lasso & Multivariate & 0.1409 & 0.2750 & 0.2969 & 0.1574 & 0.5142 \\ 
  Fixed-effects & Elastic-net & Multivariate & 0.1286 & 0.2479 & 0.2859 & 0.1359 & 0.5002 \\ 
  Fixed-effects & Elastic-net & Univariate & 0.1331 & 0.2733 & 0.3136 & \textbf{0.1368} & 0.5251 \\ 
   \hline
\end{tabular}
\label{tab:application_EXPO}
\end{table}

\subsection{Association of DNAm surrogates with CVD risk} \label{sec:cvd_risk}
\begin{figure}
    \centering
    \includegraphics[scale=.7]{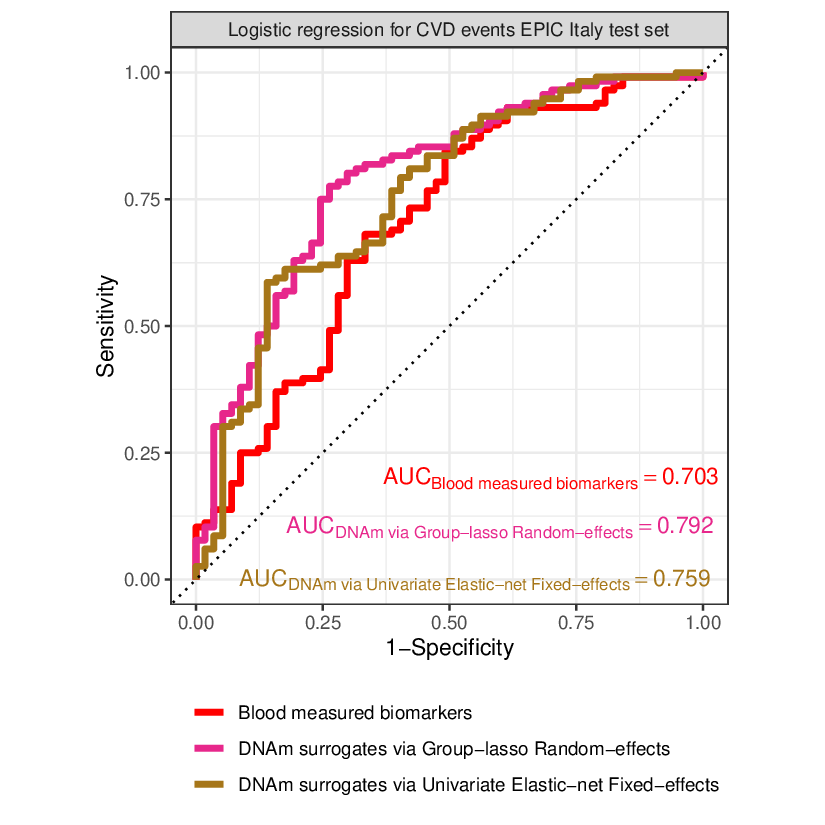}
    \caption{Receiver Operating Characteristic curves and  Area Under the Curve metrics for the association analyses of DNAm surrogates with
CVD risk, Italy EPIC test set.}
    \label{fig:cvd_risk_AUC_EPIC_test}
\end{figure}

\re{DNAm surrogates creation is not a standalone regression problem, as its primary aim is to provide reliable covariates for diseases prediction models \citep{Fernandez-Sanles2021, Odintsova2021, Hidalgo2021}. We therefore validate whether employing the estimated DNAm surrogates acts as a superior proxy of blood measured biomarkers in association analyses. In details, within the cohort of patients in the EPIC Italy test set we build logistic regression models to predict the probability of
cardiovascular risk using as regressors either the blood measured biomarkers or the two best performing DNAm surrogates devised in the previous section, adjusting for sex and age. The Receiver Operating Characteristic (ROC) curves and the associated Area Under the Curve (AUC) metrics for the considered methods are displayed in Figure \ref{fig:cvd_risk_AUC_EPIC_test}. As expected, in light of the RMSE results reported in Table \ref{tab:application}, we notice that classification performances
are similar among the competing models. Nonetheless, the logistic curves regressed on the DNAm based
surrogates seem to outperform the blood measured counterparts. Interestingly, all surrogates in Table \ref{tab:application} define logistic regression models whose AUCs are higher than those retrieved by the blood measured biomarkers, with the best performance attained by the penalized random-intercept model with multivariate group-lasso penalty.}

\re{We further assess the association of DNAm surrogates with CVD risk in the external EXPOsOMICS study. For this dataset values of the blood based biomarkers are available only for a subset of volunteers, we thus construct the logistic regression models by means of the surrogates only. Such a situation is quite common in validation data and in line with the principle DNAm surrogates were devised in the first place. Also in this context the predictive performance of our novel proposal, coupled with a multivariate group lasso penalty,  is higher with respect to state-of-the-art surrogates created via Elastic-net Fixed-effects models.} The associated Receiver Operating Characteristic curves, as well as additional figures related to the DNAm biomarkers analysis, are reported in the Supplementary Material \citep{Cappozzo2023_supplement}.

\re{The association analyses reported in this section cast light on the applicability of the devised DNAm surrogates as an enriched and patient-specific proxy of their blood measured counterparts, in both internal and external cohorts. These favorable outcomes indicate that using models based on DNAm surrogates
could be more appropriate for prediction tasks, such as CVD prevention, since they can possibly incorporate
individual characteristics not directly recorded in the blood measured biomarkers.}

\subsection{CpG sites selection and gene set enrichment analyses of inflammatory pathways} \label{sec:CpG_interpretation}

\re{In the previous sections the newly devised random-intercept model for multitask learning has demonstrated superior predictive performance when it comes to DNAm surrogates creation and CVD prediction. We hereafter examine the epidemiological rationale of the multivariate group-lasso penalty compared to univariate elastic-nets, investigating the biological reliability of the selected features (CpG sites). The univariate elastic-nets extract $178$, $504$, $518$, $79$, $497$ CpG sites for diastolic blood pressure, HDL cholesterol, LDL cholesterol, systolic blood pressure and triglycerides, respectively. As reported in Table \ref{tab:application}, the total number of unique CpGs is $1712$. However, despite the high degree of correlation among the multivariate outcomes, no CpGs were in common in the five sets, and only a minor percentage of CpGs was shared among two or more responses. Instead, as previously described , our MLMM procedure regularized with a multivariate group-lasso penalty extracts $417$ features that are associated with the five outcomes at the same time, a biological mechanism known as pleiotropy \citep{Atchley1991, Lobo2008}.}

\re{We are further interested in assessing whether the selected CpG sites are associated with specific biological pathways. To do so, gene set enrichment analyses \citep{Subramanian2005} are performed on the features retained by penalized MLMM and univariate elastic-net models. Specifically, given that the number of CpGs extracted for each method is small (hundreds of CpGs selected from an initial set of $295614$ features), an analysis based on all the biomolecular pathways described in the canonical datasets, i.e., KEGG \citep{Yi2020}, GO \citep{Harris2004} and Reactome \citep{Fabregat2018},  would be under-powered. Therefore, to overcome this limitation and with inflammation  being the main mechanism involved in the onset of the majority of chronic diseases, we focus our analyses on the $17$ inflammatory pathways described in \citet{Loza2007}.  Enrichment analyses results are summarized in Table \ref{tab:pathway_analysis}. For each list of CpGs we test for over-representation of features in inflammatory pathways using the method implemented in the \texttt{missMethyl R} package \citep{Phipson2016b}. Considering the CpGs extracted by our multivariate approach, we find significant enrichment for CpGs in four inflammatory pathways. These results agree with previous literature suggesting that hypertension and hyperlipidaemia are associated with the dysregulation of molecular pathways regulating apoptosis, oxidative stress, and the immune system \citep{Senoner2019, Dong2020}. Instead, modeling the outcomes one by one using univariate models leads to a less consistent pattern of associations. In fact, we find only one (and always different) significant pathway per analysis.}

\re{All in all, the results reported in this section support the advantages of modeling multiple correlated outcomes not only from a prediction perspective, for both blood measured biomarkers and endpoint of interest, but also considering the biological reliability of the extracted features.}

\begin{table}[]
\centering
\caption{Empirical p-values of the enrichment analyses computed using a permutation procedure via the \texttt{gometh} function in the \texttt{missMethyl R} package. Empirical p-values lower than 0.05, highlighted in bold, indicate significant overrepresentation.} 
\hspace*{-3cm}\begin{tabular}{ccccccc}
\hline
\textbf{Inflammatory pathway}        & \textbf{Group-lasso Random-effects}                   & \multicolumn{5}{c}{\textbf{Univariate Elastic-net Fixed-effects}}                                                                                                                                     \\
\textbf{}                            & \textbf{}                              & \textbf{DBP} & \textbf{HDL}                           & \textbf{LDL}                            & \textbf{SBP}                          & \textbf{TG}                            \\
\hline
Leukocyte signaling                  & \textbf{0.006} & 0.14         & 0.35                                   & 0.01                                    & 1                                     & 1                                      \\
ROS/Glutathione/Cytotoxic granules   & \textbf{0.009} & 0.06         & 0.15                                   & 1                                       & \textbf{0.03} & 0.15                                   \\
Apoptosis Signaling                  & \textbf{0.01}  & 1            & 1                                      & 0.17                                    & 1                                     & 0.07                                   \\
Natural Killer Cell Signaling        & \textbf{0.01}  & 0.36         & 0.10                                   & 1                                       & 1                                     & 1                                      \\
PI3K/AKT Signaling                   & 0.18                                   & 1            & 0.26                                   & 0.03                                    & 1                                     & 0.22                                   \\
Innate pathogen detection            & 0.26                                   & 0.12         & 0.30                                   & 0.31                                    & 1                                     & 1                                      \\
Cytokine signaling                   & 0.36                                   & 1            & 1                                      & \textbf{0.0003} & 1                                     & 0.10                                   \\
Adhesion-Extravasation-Migration     & 1                                      & 0.18         & \textbf{0.003} & 0.02                                    & 0.09                                  & 0.10                                   \\
Calcium Signaling                    & 1                                      & 1            & 1                                      & 0.08                                    & 1                                     & 1                                      \\
Complement Cascase                   & 1                                      & 1            & 1                                      & 1                                       & 1                                     & 0.16                                   \\
Glucocorticoid/PPAR signaling        & 1                                      & 1            & 0.10                                   & 1                                       & 0.17                                  & 0.10                                   \\
G-Protein Coupled Receptor Signaling & 1                                      & 1            & 1                                      & 1                                       & 0.05                                  & 0.27                                   \\
MAPK signaling                       & 1                                      & 1            & 0.14                                   & 0.49                                    & 1                                     & \textbf{0.005} \\
NF-kB signaling                      & 1                                      & 1            & 0.16                                   & 0.16                                    & 1                                     & 0.15                                   \\
Phagocytosis-Ag presentation         & 1                                      & 1            & 1                                      & 0.26                                    & 1                                     & 1                                      \\
Eicosanoid Signaling                 & 1                                      & 1            & 1                                      & 1                                       & 1                                     & 1                                      \\
TNF Superfamily Signaling            & 1                                      & 1            & 0.01                                   & 0.14                                    & 1                                     & 1                                     \\
\hline
\end{tabular}
    \label{tab:pathway_analysis}
\end{table}

\section{Discussion and further work} \label{sec:conclusion}
In the present paper we have proposed a novel framework for mixed-effects multitask learning suitable for high-dimensional data. The ubiquitous presence in modern applications of ``$p$ bigger than $N$'' problems asks for the development of ad-hoc statistical tools able to cope with such scenarios. By resorting to penalized likelihood estimation, we have devised a general purpose EM algorithm capable of accommodating any penalty type that has been previously defined for fixed-effects models. We have examined three functional forms for the penalty term, discussing pros and cons of each and providing convenient routines for model fitting. The proposal has been accompanied by some considerations on distinguishing features, like how to quantify response specific random-effects, and other more general issues concerning initialization, convergence and model selection.

The work has been motivated by the problem of developing a multivariate DNAm biomarker of cardiovascular and high blood pressure comorbidities from a multi-center study.  The EPIC Italy dataset has been analyzed using Diastolic Blood Pressure, Systolic Blood Pressure,  High Density Lipoprotein, Low Density Lipoprotein and Triglycerides as response variables, regressing them on $62128$ CpG sites and accounting for between-center heterogeneity. Our modeling framework, coupled with a multivariate group-lasso penalty, has demonstrated to outperform the state-of-the-art alternative, both in terms of predictive power and biomedical interpretation. Remarkably, the number of CpG sites deemed as relevant in the multi-dimensional surrogate creation was found to be lower than those identified by separately fitting penalized models for each risk factor. Decreasing the amount of relevant CpG sites is crucial to reduce sequencing costs for future studies, with the final aim of querying only a limited number of targeted genomic regions. Such a result may thereupon favor the adoption of our methodological approach for building DNAm surrogates.

The devised pipeline possesses also some limitations. The EWAS results are adjusted for clinical covariates external to the analysis, and this may thus affect the pre-processing outcome. Moreover, the level of strictness in the screening process is influenced by the chosen threshold on the p-values. On this wise two different, yet both sensible, strategies can be adopted. On the one hand, one may rely on the ``Occam's razor'' principle, preferring to use a stricter threshold being it the simplest and fastest option. On the other hand, one can positively include many redundant variables in the design matrix, relying on the model ability to shrink coefficients of irrelevant features to zero. Concurrently, further insights about the associations between DNA methylation and blood-measured biomarkers may be unraveled by means of sparse multiple canonical correlation analysis \citep{Witten2009,Witten2009a,Rodosthenous2020}; while other modeling approaches such as deep-learning \citep{Nguyen2022, Yuan2022.09.02.22279533}, Bayesian methods \citep{Zhao2021a, Zhao2021} and boosting machines \citep{Sigrist2022} could be profitably adapted to build DNAm multidimensional surrogates.

A direction for future  research concerns promoting the application of the proposed procedure in creating additional multi-dimensional DNAm biomarkers, conveniently embedding mixed-effects and customized penalty types. \re{In this regard, of particular interest may be the definition of a shrinkage term for which the grouping in $\boldsymbol{B}$ is introduced from both responses and predictors: the former is induced by the multivariate nature of $\boldsymbol{Y}$ (i.e., the $r$ responses), while the latter can stem from any structure present in $\boldsymbol{X}$ (e.g., CpG islands). Such a problem could be solved by extending the Multivariate Sparse Group Lasso, proposed by \cite{Li2015}, to the mixed-effects framework.}

In addition, having assumed random intercepts for each and every component in a low-dimensional response framework was only motivated by the application at hand, and it may not be valid in general. Thus, a two-fold methodological development naturally arises: a first one concerning the definition of  response-specific random-effects in multitask learning and another accounting for the inclusion of custom penalties when dealing with high-dimensional response variables. Furthermore, the latter may also possess a mixed-type structure, with components simultaneously be nominal, ordinal, discrete and/or continuous. Some proposals are currently under study and they will be the object of future work.

\bibliography{arxiv_library_cappozzo_ieva_fiorito}       

\end{document}